\documentclass[aps,reprint,floatfix,footinbib]{revtex4-1}
\usepackage{float}
\usepackage{epsfig}
\usepackage{graphicx}
\usepackage{amsmath}
\usepackage{amsfonts}
\usepackage{amssymb}
\usepackage{bm}
\usepackage{bbold}
\usepackage{epsfig}
\usepackage{graphicx}
\usepackage{color}
\usepackage{pictex}
\usepackage{mathtools}
\usepackage{extarrows}
\usepackage{footnote}
\usepackage{footmisc}
\newcommand{\s}{\scriptscriptstyle}
\begin{document}
\title{Disorder lines, modulation,  and partition function zeros in free fermion models}
\author{P. N. Timonin}
\affiliation{Physics Research Institute,
Southern Federal University, 194 Stachki ave.,  Rostov-on-Don, 344090 Russia}
\author{Gennady Y. Chitov}
\affiliation{Department of Physics, Laurentian University, Sudbury, Ontario,
P3E 2C6 Canada}
\date{\today}

%
%
\begin{abstract}
The modulation is analyzed from the analytical properties of zeros of free fermionic partition function on the complex plane of wave numbers.
It is shown how these properties are related to the oscillations of correlation functions. This approach can be used for analysis of phase transitions
with local or nonlocal order parameters, as well as for the disorder lines.
We find an infinite cascade of disorder lines at finite temperature in the quantum $XY$ chain (equivalent to free fermions).
The well-known ground state factorization on the disorder line, and consequently, disentanglement, is shown
to follow directly from analytical properties of this model on the complex plane.
From the quantum-classical correspondence the results for the chain are used to detect the disorder lines in several frustrated 2D Ising models. The present formalism can be applied to other fermionic models in two and three spatial dimensions. In particular, we find the temperature-dependent Fermi wave vector of oscillations in the degenerate gas of 3D fermions, which naturally leads in the limit $T \to 0$ to the definition of the Fermi energy as the surface of quantum criticality. The modulation is a very common phenomenon, and it occurs in a large variety of models. The important point is that all these modulation transitions can be related to the complex zeros of partition functions, as done in the present study.

\end{abstract}
\maketitle

%
%
%
\section{Introduction}\label{Intro}
%
%
%
%
The fundamental notions of the Landau paradigm are local order parameter and the symmetry it breaks
spontaneously  \cite{LandauV5}. There has been a huge recent effort to understand whether
various low-dimensional fermionic or spin systems as quantum spin liquids,
frustrated magnetics, topological and Mott insulators, etc \cite{Fradkin:2013,Bernevig:2013,RyuSchnyder:2010,Montorsi:2012}, which lack
conventional long-ranged order even at zero temperature, can be dealt within the Landau framework, or a new
paradigm of topological order \cite{Wen:2017} needs to be used instead.

The Landau paradigm, although extended to incorporate nonlocal string order \cite{denNijs:1989} and hidden symmetry breaking
\cite{Oshikawa:1992,*Kennedy:1992,*Kohmoto:1992}, remains instrumental even for nonconventional orders \cite{Chitov:2017JSM,Chitov:2018,Chitov:2019}.
The local and nonlocal string order parameters in the extended formalism are related by duality, and
probing a phase transition and relevant order becomes a matter of appropriate choice of variables
\cite{ChenHu:2007,Xiang:2007,NussinovChen:2008,Nussinov:2013,Smacchia:2011,Chitov:2017JSM,Chitov:2018,Chitov:2019}.

Probably the most fundamental rigorous approach in the theory of phase transitions, applicable whatever is the nature of order parameter or
symmetry breaking,  was pioneered by Yang and Lee  \cite{YangLee:1952,*LeeYang:1952}. They related transitions to the zeros of model's partition function, which in the zero temperature limit becomes the requirement of gap closure. The original analysis of Yang and Lee of the ferromagnetic Ising model
was further extended for other models and the cases out of equilibrium. For a short list of references, see, e.g.
\cite{Fisher:1965,Fisher:1980,Matveev:2008,Bena:2005,Wei:2014}, and more references in there.

In 1970 Stephenson \cite{Stephenson-I:1970,Stephenson-II:1970,Stephenson:1970PRB} found a new type of weak transitions in classical Ising models which he dubbed ``disorder lines" (DL). The transition consists in changing the behavior of the correlation functions from monotonic exponential decay to the exponential decay modulated by incommensurate oscillations. The weakness of such transition is manifested in the behavior of the correlation length, which demonstrates
only a cusp at the disorder line  point. A similar transition was later found in the $XY$ quantum chain by Barouch and McCoy \cite{McCoyII:1971}.
Disorder lines, or more broadly, modulation transitions are quite general phenomena occurring in a large variety of models  \cite{Nussinov:2011,Nussinov:2012,Salinas:2012}, including the recently reported pattern formation in QCD \cite{Ogilvie:2020}.

An important conclusion of our earlier work on the classical Ising chain \cite{Chitov:2017PRE} is that the disorder lines found by Stephenson, and
moreover, the infinite cascades of disorder lines found in \cite{Chitov:2017PRE}, are zeros of the partition function in the range of complex magnetic field. Similarly, the appearance of modulations in the ground state of the quantum $XY$ chain \cite{McCoyII:1971}
is related to zeros of model’s spectrum on the complex plane of wave vectors $k \in \mathbb{C}$ \cite{Franchini:2017}.

In this work we propose a unifying framework based on the analysis of the roots for zeros of the partition function on the complex plane of wave numbers.
These roots combine all possible solutions corresponding to the continuous phase transitions, as well as to disorder lines where the modulation sets in.
From the analytical properties of the two-point Majorana correlation functions on the complex plane, we relate the appearance of oscillations in those functions on the disorder lines to the analytical properties of the complex roots of the partition function. We mainly discuss our results in the context of the simple quantum $XY$ chain in transverse field which is dually equivalent to free fermions. Since the transfer matrices of several 2D Ising models
commute with the Hamiltonian of quantum spin chain at some special points \cite{Suzuki-I:1971,*Suzuki:1971,Krinsky:1972,Stephen:1972,Peschel:1981,Peschel:1982,Rujan-II:1982}, we extend our analysis on the disorder lines in the 2D Ising models. The present formalism can be straightforwardly applied for tight binding lattice fermions or Fermi gas in two and three spatial dimensions.
In particular, we show that the complex roots for the zeros of the partition function of the 3D non-relativistic degenerate gas of fermions, naturally lead to the definition of temperature-dependent Fermi wave vector. The latter defines the gapless Fermi surface of quantum criticality in the limit $T \to 0$.

The rest of the paper is organized as follows:
In Sec.~\ref{DLIntro} we present some general results for an arbitrary model of non-interacting 1D lattice fermions to possess disorder lines.
In Sec.~\ref{XYchain} we take $XY$ quantum chain in transverse field as an example to present our main results on zeros of
partition functions, modulation and cascades of disorder lines at finite temperatures, factorization of the ground state
and disentanglement.   Sec.~\ref{Ising} presents the results for disorder lines in several 2D Ising model based on the equivalence
between quantum and classical models. In Sec.~\ref{3DFG} we extend our analysis for the complex zeros of the partition function of
the degenerate 3D Fermi gas with a Fermi surface.  The results are summarized in the concluding Sec.~\ref{Concl}.
%
%
%

%
%
%
%
\section{Disorder lines in quantum chains: General analysis}\label{DLIntro}
%
%
%
%
We consider non-interacting spinless fermions $c_n$ defined on the sites
of a chain. The model is assumed to be periodic with the period of one lattice spacing,
and in the reciprocal space its generic Hamiltonian can be written as \cite{LiebSM:1961}
\begin{equation}
\label{Hspinor}
 H= \frac12 \sum_{k}\psi^{\dag}_{k}\hat{\mathcal{H}}(k) \psi_{k}~,
\end{equation}
where the fermions are unified in the spinor
\begin{equation}
  \psi_{k}^{\dag}=\left(c^{\dag}(k),c(-k) \right)~,
\label{spinor1}
\end{equation}
with the wave numbers restricted to the Brillouin zone $k \in
[-\pi,\pi]$, and we set the lattice spacing $a=1$.
We choose the $2\times 2$ Hamiltonian  matrix in the general form \cite{LiebSM:1961}:
\begin{equation}
\label{Hk}
  \hat{\mathcal{H}}(k) = \left(%
\begin{array}{cc}
  A & B \\
  B^\ast  & -A \\
\end{array}%
\right)~,
\end{equation}
with $A(k) \in \mathbb{R}$ and $B(k)=|B|e^{i \varphi}$.
We diagonalize \eqref{Hspinor} by the unitary Bogoliubov transformation
\begin{equation}
\label{UH}
  \hat{U}\hat{\mathcal{H}}\hat{U}^\dag =\varepsilon(k) \hat{\sigma }^z
\end{equation}
with
\begin{equation}
\label{U}
   \hat U= \left(
                   \begin{array}{cc}
                     \cos \vartheta  & -\sin \vartheta e^{i \varphi}\\
                      \sin \vartheta &  \cos \vartheta  e^{i \varphi}\\
                   \end{array}
                 \right)~.
\end{equation}
The Bogoliubov angle $\vartheta$ is defined by the following equation:
\begin{equation}
\label{ThetaB}
  \tan \vartheta = \sqrt{\frac{\varepsilon-A}{\varepsilon+A}}=\frac{|B|}{\varepsilon+A}~,
\end{equation}
where $\varepsilon(k)=\sqrt{A^2+|B|^2}$ is the spectrum of the Hamiltonian.

The new fermionic operators in the diagonalized representation are related to the original fermions as
\begin{equation}
\label{etaU}
  \left(
    \begin{array}{cc}
      \eta(k) \\
      \eta^\dag(-k) \\
    \end{array}
  \right)=
  \hat{U} \psi_k
\end{equation}
In this non-interacting model all correlation functions can be expressed via two-point average of Majorana
operators \cite{LiebSM:1961,Franchini:2017}
\begin{equation}
\label{G}
G_{r} =\left\langle i b_{l} a_{l+r} \right\rangle =
\int _{-\pi }^{\pi }\frac{dk}{2\pi }  e^{ikr} G(k),
\end{equation}
where
\begin{equation}
\label{GkDef}
  G(k)=\left\langle i b(-k)a(k) \right\rangle
\end{equation}
The original lattice fermion is represented via two self-adjoint (Majorana) operators as:
\begin{equation}
\label{Maj}
   a_n +i b_n  \equiv 2 c^{\dag}_n~.
\end{equation}
In terms of the Fourier transforms:
\begin{eqnarray}
\label{ABk}
  a(k) &=& c(k)+c^\dag(-k), \nonumber \\
  ib(k) &=& c(k)-c^\dag(-k)
\end{eqnarray}
with
\begin{equation}
\label{akbk}
  a^\dag(k)=a(-k),~ ~ b^\dag(k)=b(-k)~.
\end{equation}
From the above equations one readily finds the Fourier transform of the Majorana correlation function as
\begin{eqnarray}
\label{GDA}
  G(k) &=&  D(k)
  \left[\left\langle \eta^\dag (k) \eta(k) \right\rangle -\left\langle \eta(-k) \eta^\dag (-k) \right\rangle \right]
  \nonumber \\
  &=& D(k) \tanh \frac{\varepsilon (k) }{2T}~,
\end{eqnarray}
where the generating function $D(k)$ is found from the components of the unitary Bogoliubov matrix:
\begin{equation}
\label{Dk}
  D(k)=\left(U_{11} -U_{12} \right)\left(U_{11}^{*} +U_{12}^{*} \right)=
  \frac{A-(B-B^*)/2}{\sqrt{A^2+|B|^2}}~.
\end{equation}
Introducing the complex variable
\begin{equation}
\label{z}
z=e^{i k}
\end{equation}
we can unify the above results in a single expression for the Majorana correlation function
as a loop integral on the complex plane $z$:
\begin{equation}
\label{Loop}
G_{r}  =\oint _{\left|z\right|=1} \frac{dz}{2\pi i}
         z^{r-1} D\left(z\right)\tanh \frac{\varepsilon \left(z\right)}{2T}~.
\end{equation}
When $r>1$ the contributions to $G_r$ in Eq.~\eqref{Loop} come from nonanalyticies of $G(z)$ inside the unit circle. Except for a few particular limits
to be discussed later, the analytical continuation of the generating function on the complex plane $D(z)$ has branch cuts, while
$\tanh \frac{\varepsilon \left(z\right)}{2T}$ has poles at
\begin{equation}
\label{Poles}
\varepsilon \left(z\right)=i \omega_n~, ~~\omega_n \equiv \pi T\left(2n+1\right)
\end{equation}
(we set $\hbar=k_B=1$). As shown below, the appearance  of the poles \eqref{Poles} inside the unit circle signals the onset the incommensurate (IC) oscillations in $G_r$, i.e, the \textit{disorder line} as defined by Stephenson \cite{Stephenson-I:1970,Stephenson:1970PRB}.

A very fundamental point is that equation \eqref{Poles} defines \textit{zeros of the partition function} of a free-fermionic system  \cite{TongLiu:2006}.
When the magnetic field is analytically continued on the complex plane, such zeros are called the Lee-Yang zeros \cite{YangLee:1952,*LeeYang:1952},
while in case of complex temperature they are called the Fisher zeros \cite{Fisher:1965}. Disorder lines are zeros of the partition function in the complex
range of parameters, as we have shown in earlier work on the classical Ising and quantum fermionic chains  \cite{Chitov:2017PRE,Chitov:2018}.
In the present work we systematically identify and analyse the disorder lines as zeros of the partition function  in the range of complex wave numbers.

Denoting the roots of the partition function by $\Lambda_\alpha (n)$, one can write equation \eqref{Poles} as
\begin{equation}
\label{ZeroDecomp}
 \varepsilon^2(z) + \omega_n^2 = \mathcal{A}\prod_{\alpha=1}^m  \left(z-\Lambda_\alpha (n) \right)~.
\end{equation}
Use it in the expansion
\begin{equation}
\label{th}
  \tanh \frac{\varepsilon}{2T} = 4 T  \sum _{n=0}^{\infty} \frac{\varepsilon}{\varepsilon^2+\omega_n^2}
\end{equation}
for the Majorana correlation function  in equation \eqref{Loop} yields:
\begin{widetext}
\begin{equation}
\label{GnDecomp}
G_{r}  = \oint _{\left|z\right|=1} \frac{dz}{2\pi i}
         z^{r-1} D\left(z\right)
         \varepsilon (z)\sum _{n=0}^{\infty} \frac{4 T}{\mathcal{A}}  \prod_{\alpha=1}^m  \left(z-\Lambda_\alpha (n) \right)^{-1}
         ~.
\end{equation}
\end{widetext}
Any complex pole of \eqref{GnDecomp} inside the unit circle $\Lambda_\alpha (n) =e^{iq_n-\kappa_n}$ gives the contribution
\begin{equation}
\label{dG}
\delta G_{r+1}^{(n)} \propto   e^{-\kappa_n r} e^{iq_n r} ~
\end{equation}
of an oscillating decaying mode into $G_{r}$. Since there exists an infinite set of the oscillating modes \eqref{dG} with different $n$,
the leading asymptotic behavior  $r \gg 1 $ (whether $q_n =0$ or $q_n \neq 0$) is determined by the minimal $\kappa_n$ which thus determines
the inverse correlation length. For instance for the case of the $XY$ chain discussed below the minimal $\kappa_n$ occurs at $n=0$, but we do not have
a general proof that in some models it can not happen for another $n>0$.
These oscillations manifests themselves in the behavior of other correlation functions: the latter are expressed via determinants of the Toeplitz matrices where $G_{r}$ are the elements of those matrices \cite{LiebSM:1961}, resulting so to leading order to the exponentially decaying oscillations. \cite{McCoyII:1971}

\begin{figure}[h]
\centering{\includegraphics[width=6.5 cm]{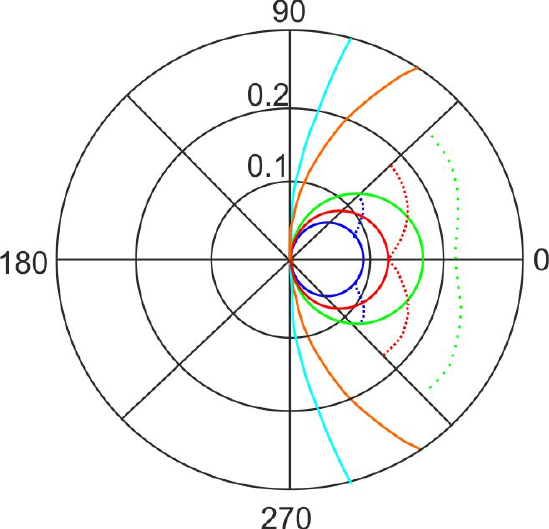}}
 \caption{Contours  $C_{\s R}(\mathbf{v})=C_{\s R}(h,\gamma,k)$ (solid lines) for the XY chain at $\zeta \equiv h/(1-\gamma^2)=0.5,1$ (open) and $\zeta=3,4,5$ (closed) plotted on the complex plane $z$. The radius of the outer circle is 0.3. The contours $C_{\s I}(h,\gamma,k)$, $n = 0$ (dotted lines)
 are shown for three qualitatively different cases: (1), green, $\zeta=3$ ($\gamma=0.5,~h=2.25$) and $T<T_{\s DL,0}$ ($T=0.35,~T_{\s DL,0}=0.382$), no intersections of $C_{\s R}$ and $C_{\s I}$; (2), red,  $\zeta=4$ ($\gamma=0.5,~h=3$), single intersection on the real axis at the critical point $T=T_{\s DL,0}=0.582$; (3), blue, $\zeta=5$ ($\gamma=0.5,~h=3.75$) and $T>T_{\s DL,0}$ ($T=0.8,~T_{\s DL,0}=0.671$),  two intersections of $C_{\s R}$ and
 $C_{\s I}$ yield the wave number of oscillations $q_\circ(\zeta=5,T=0.8) \approx 0.31$ in accordance with available analytical results.}
 \label{Contours}
\end{figure}

There is also a more pictorial way to analyse general properties of the disorder line solutions of \eqref{Poles} without choosing a particular model.
We can rewrite \eqref{Poles} as a couple of equations for the real and imaginary parts of $\varepsilon$:
\begin{eqnarray}
  \varepsilon ' (z, \mathbf{v})&=&0
  \label{epsRe} \\
   \varepsilon '' (z, \mathbf{v})&=& (2n+1)\pi T~,
  \label{epsIm}
\end{eqnarray}
where $\mathbf{v}$ stands as a shorthand for the Hamiltonian's  parameters not  shown explicitly.
The solutions of the first equation \eqref{epsRe} can be depicted as some contours $C_{\s R}(\mathbf{v})$ on the complex plane.
Similarly, the solutions of the second equation \eqref{epsIm} define another set of contours denoted as $C_{\s I}(\mathbf{v})$.
The oscillations of $G_r$ discussed above occur when within a certain range of parameters $\mathbf{v}_\circ \in \mathbf{v}$
two contours $C_{\s R}(\mathbf{v})$ and $C_{\s I}(\mathbf{v})$ intersect inside a unit circle on the complex plane $z$.
If such intersections are impossible, the model does not have disorder lines.
As an example we present in Fig.~\ref{Contours} the graphical solution of Eqs. \eqref{epsRe} and \eqref{epsIm} for the $XY$ chain.

A salient point we infer from equations \eqref{Poles},\eqref{ZeroDecomp},\eqref{epsRe}, and \eqref{epsIm} is the existence of an infinite sequence of the disorder line temperatures
\begin{equation}
\label{TDLgen}
T_{\s DL,n} =\frac{\varepsilon '' \left(\Lambda_\alpha (n),\mathbf{v}_\circ \right)}{\left(2n+1\right)\pi } \equiv \frac{T_{\s DL,0} }{2n+1}
\end{equation}
which control appearance of the oscillations $\propto q_n$ of the correlation function $G_r$.

The fermion spectrum analytically continued onto the complex plane $\varepsilon \left(z\right)$ contains the information on the existence
of disorder lines in a given model. If they exist, one can find from $\varepsilon \left(z\right)$ the disorder line temperatures, the wave numbers of oscillations, and the correlation lengths, as functions of temperature and parameters of the Hamiltonian.

%
%
%
%
%
\section{Applications to $XY$ chain}\label{XYchain}
%
%
%
%
%
%
%
\subsection{Zeros of partition function}\label{LYZGen}
%
%
%
As probably the simplest albeit non-trivial model to explain the salient points of our analysis, we
take the quantum $XY$ chain in transverse magnetic field:
\begin{equation}
\label{XYHam}
   H =-\sum_{n=1}^{N} \Big\{ \frac{J}{4}  \big[ (1+\gamma )
 \sigma_{n}^{x}\sigma_{n+1}^{x}
 + (1-\gamma ) \sigma_{n}^{y}\sigma_{n+1}^{y} \big]
 + \frac12 h \sigma_{n}^{z} \Big\}~.
\end{equation}
Here $\sigma$-s are the standard Pauli matrices, coupling $J>0$ is ferromagnetic.
We assume $0<\gamma \leq 1$. The range of negative $\gamma$ is readily available from
model's symmetry under exchange $\gamma \leftrightarrow -\gamma$ and $x \leftrightarrow y$.
The material presented in this section is quite well known \cite{McCoy:2010,Franchini:2017}, in particular the equations for the DL
at finite temperature were reported in \cite{McCoyII:1971}. However Barouch and McCoy did not elaborate on their
findings to analyze the DLs in more depth.\textit{ The novelty of the present analysis of the $XY$ chain is to advance a common framework
unifying DLs and conventional phase transitions as different types of zeros of the model's partition function.}

The Jordan-Wigner (JW) transformation \cite{LiebSM:1961,Franchini:2017} maps \eqref{XYHam} onto the free-fermionic
Hamiltonian
\begin{equation}
\label{XYFermi}
   H =-\sum_{n=1}^{N}~ \Big\{ \frac{J}{2}  \big[ c_{n}^\dag c_{n+1} +
   \gamma c_{n}^\dag c_{n+1}^\dag +\mathrm{h.c.} \big]
             +   h \Big( c_{n}^\dag c_{n}- \frac12~ \Big) \Big\} ~,
\end{equation}
The  zeros of the model's partition function are determined by the following equation \cite{TongLiu:2006}
\begin{equation}
\label{LYZdef}
 \varepsilon(k)=
 \sqrt{( h -\cos k )^2+ \gamma^2 \sin^2 k}= i \omega_n~,
\end{equation}
From now on we set the units such that $J=1$.
Using the complex variable \eqref{z}, the spectrum can written as
\begin{equation}
\label{epsZ}
 \varepsilon^2(z)=\frac{(1+\gamma)^2}{4}
 (z-\lambda_+) (z-\lambda_-)(z^{-1} -\lambda_+)(z^{-1}-\lambda_-)~,
\end{equation}
where
\begin{equation}
\label{lampm}
\lambda_\pm = \frac{h \pm \sqrt{h^2+\gamma^2-1}}{1+\gamma}~.
\end{equation}
The generating function \eqref{Dk} for this model reads \cite{McCoy:2010}
\begin{equation}
\label{DzXY}
   D(z)
   = \left[ \frac{(z-\lambda_+)(z-\lambda_-)
   }{(1-z \lambda_+)(1-z \lambda_-)}  \right]^{1/2}
\end{equation}
Equation \eqref{LYZdef} for the partition function zeros
is equivalent to
\begin{eqnarray}
\label{LYZerosXY}
 && \varepsilon^2(z) + \omega_n^2 =  \frac{1-\gamma^2}{4\Lambda_+ \Lambda_- }  \nonumber \\
 &\times & (z-\Lambda_+) (z-\Lambda_-)(z^{-1} -\Lambda_+)(z^{-1}-\Lambda_-),
\end{eqnarray}
with the roots
\begin{equation}
\label{Lampm}
\Lambda_\pm = \frac{h \pm R -\sqrt{(h \pm R)^2-(1-\gamma^2)^2}}{1-\gamma^2}~,
\end{equation}
where
\begin{equation}
\label{R}
 R \equiv \sqrt{\gamma^2(h^2+\gamma^2-1)-(1-\gamma^2)\omega_n^2}~.
\end{equation}
Note that in the limit $T \to 0:$ $\Lambda_\pm \to \lambda_\pm$.

\begin{widetext}
The Majorana correlation function
\begin{equation}
\label{GnSeries}
G_{r}  = \frac{16 T}{1-\gamma^2}  \oint _{\left|z\right|=1} \frac{dz}{2\pi i}
         z^{r-1} D\left(z\right)
         \varepsilon (z)\sum _{n=0}^{\infty} \frac{\Lambda_+ \Lambda_-}{(z-\Lambda_+) (z-\Lambda_-)(z^{-1} -\Lambda_+)(z^{-1}-\Lambda_-)}
         ~.
\end{equation}
\end{widetext}
One needs to keep in mind that the roots $\Lambda_\pm$ in the above expressions depend on $n$.

Conventional phase transitions correspond to the zeros of partition function occurring for real wave numbers $k$ in \eqref{z}, i.e.,
for $z$ lying on the unit circle, that is
\begin{equation}
\label{Lam1}
\left| \Lambda_\pm \right|= 1~.
\end{equation}
One can show that $\Lambda_\pm \in \mathbb{R}$ and $\Lambda_\pm \leq 1$ when $R$ defined by \eqref{R} is real, i.e., $R\in \mathbb{R}$.
The condition for the bigger root to reach unity is:
\begin{equation}
\label{PTcond}
\Lambda_+= 1:~~ (h \pm 1)^2+\omega_n^2=0~,
\end{equation}
which can be satisfied only at $T=0$ for two values of the external field $h=\pm 1$. These two
well-known lines of ferromagnetic-paramagnetic quantum phase transitions \cite{Franchini:2017} are shown on the phase diagram in
Fig.~\ref{PhaseDiag}.

The condition  \eqref{Lam1} can be also satisfied when $\gamma=0$ and $|h|<1$, again at zero temperature only. In this case
the roots are complex conjugate
\begin{equation}
\label{ICcond}
 \Lambda_\pm =h \pm i \sqrt{1-h^2}~.
\end{equation}
The solution \eqref{ICcond} engenders the line of quantum criticality corresponding to the gapless IC phase with the wave number $k=\arccos h$ \cite{Franchini:2017}. This IC line separates two ordered (at $T=0$ only!) phases with magnetizations $m_x$ and $m_y$, and it is also shown
on the $(h,\gamma)$ plane in Fig.~\ref{PhaseDiag}.

Two cases \eqref{PTcond} and \eqref{ICcond} corresponding to the two continuous quantum phase transitions in the $XY$ chain, exhaust
possible solutions of \eqref{LYZdef} or, equivalently of \eqref{LYZerosXY} with a real wave number
\begin{equation}
\label{kLog}
 k=-i \ln \Lambda~.
\end{equation}
Other solutions for zeros of the partition function exist at $T>0$ and at complex wave numbers
\begin{equation}
\label{kC}
 k \equiv q+i \kappa, ~z=e^{iq -\kappa}
\end{equation}
They correspond to disorder lines which can be thought of as ``weak transitions", analyzed in the following subsection.

%
%
%
\subsection{Disorder lines at finite temperature }\label{DLs}
%
%
%
Similarly to the classical Ising chain \cite{Chitov:2017PRE}, the $XY$ model in transverse field  possesses an infinite sequence of
disorder lines (weak thermal transitions).

The transition between the regimes of monotonous and oscillating decay of correlation functions, i.e., a disorder line,
can occur only when the roots  $\Lambda_\pm$ from real become complex.  One can check that this happens when the expression under the
radical in \eqref{R} changes its sign from positive to negative. It is possible in principle only if
\begin{equation}
\label{MinCond}
 \underline{\mathbf{(I):}~~~h^2+\gamma^2>1}
\end{equation}
which defines the boundary on the plane $(h,\gamma)$ of model's parameters where the disorder lines can occur.
If the above condition is satisfied, we can use \eqref{R} to define
\begin{equation}
\label{Tdl}
 T_{\s DL,n}= \frac{T_{\s DL,0} }{2n+1},~~T_{\s DL,0} \equiv \frac{\gamma}{\pi} \sqrt{\frac{h^2 +\gamma^2 -1}{1-\gamma^2 }}~.
\end{equation}
At
\begin{equation}
\label{Tbelow}
 \underline{\mathbf{(Ia):}~~~T<T_{\s DL,n}}~\longmapsto~\Lambda_\pm \in \mathbb{R}~,
\end{equation}
no oscillations due to the $n-$th root. For the root parameters \eqref{kC} we find
\begin{equation}
\label{CosReal}
 \cos k = \frac{z+z^{-1}}{2}=\frac{h}{1-\gamma^2} \pm \pi(2n+1) \sqrt{\frac{T_{\s DL,n}^2-T^2}{1-\gamma^2}}
\end{equation}
and
\begin{eqnarray}
  q &=& 0~, \\
\label{q1}
 \cosh \kappa &=&  \frac{h}{1-\gamma^2} \pm \pi(2n+1) \sqrt{\frac{T_{\s DL,n}^2-T^2}{1-\gamma^2}}~.
\label{kap1}
\end{eqnarray}
One can check that $\kappa>0$, so $\Lambda_\pm <1$.

In the temperature range
\begin{equation}
\label{Tbelow2}
 \underline{\mathbf{(Ib):}~~~T>T_{\s DL,n}}~\longmapsto~\Lambda_\pm \in \mathbb{C}~,
\end{equation}
there are oscillations due to the $n$-th root which set in at the critical temperature $T_{\s DL,n}$. The parameters of the
root are found from
\begin{eqnarray}
\label{CosCompl}
  \cos k &=& \frac{z+z^{-1}}{2}=  \cos q \cosh \kappa -i \sin q \sinh \kappa \nonumber \\
  &=& \frac{h}{1-\gamma^2} \pm i \pi(2n+1) \sqrt{\frac{T^2-T_{\s DL,n}^2}{1-\gamma^2}}~,
\end{eqnarray}
whence the IC wave numbers of oscillations $\pm q$ and the imaginary part of the wave number $\kappa$ are determined by the
following parametrization relations:
\begin{eqnarray}
  \cos^2 q &=& Q-\sqrt{Q^2-\frac{h^2}{(1-\gamma^2)^2}}   ~, \\
\label{q2}
 \cosh \kappa &=&  \frac{h}{(1-\gamma^2)\cos q} ~,
\label{kap2}
\end{eqnarray}
where we introduced the auxiliary parameter
\begin{equation}
\label{Q}
  Q \equiv \frac12 \left(1+ \frac{h^2+|R|^2}{(1-\gamma^2)^2} \right)~.
\end{equation}
One can show that in the range \eqref{MinCond} and \eqref{Tbelow} the above relations yield $\cos^2 q <1$ and $\cosh \kappa>1$.

Inside the circle
\begin{equation}
\label{MinCond2}
 \underline{\mathbf{(II):}~~~h^2+\gamma^2<1}
\end{equation}
no disorder lines exist, since the roots $\Lambda_\pm$ are always complex, i.e., the IC oscillations are present at arbitrary temperature.

\begin{figure}[h]
\centering{\includegraphics[width=8.5cm]{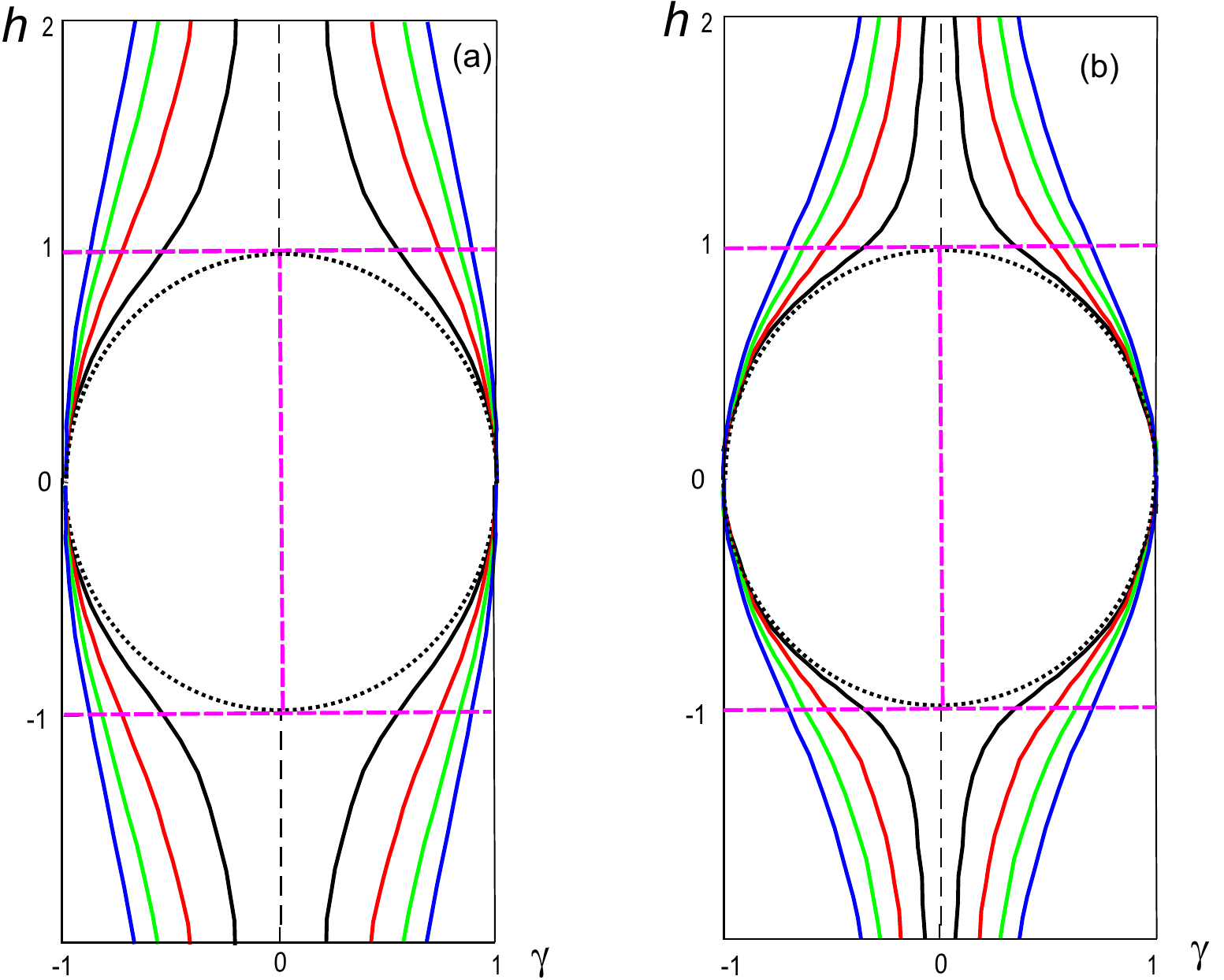}}
 \caption{(Disorder lines $h_{\s DL} (\gamma ,T,n)$ of the XY chain in the $(h,\gamma)$-plane
 for n =1 (black), 3 (red), 5 (green), 7 (blue) at T=0.015 (a) and T=0.035 (b).
 Dotted line is the $T \to 0$ limit $h^2 +\gamma^2 =1$.}
 \label{PhaseDiag}
\end{figure}

\begin{figure}[h]
\centering{\includegraphics[width=8.0 cm]{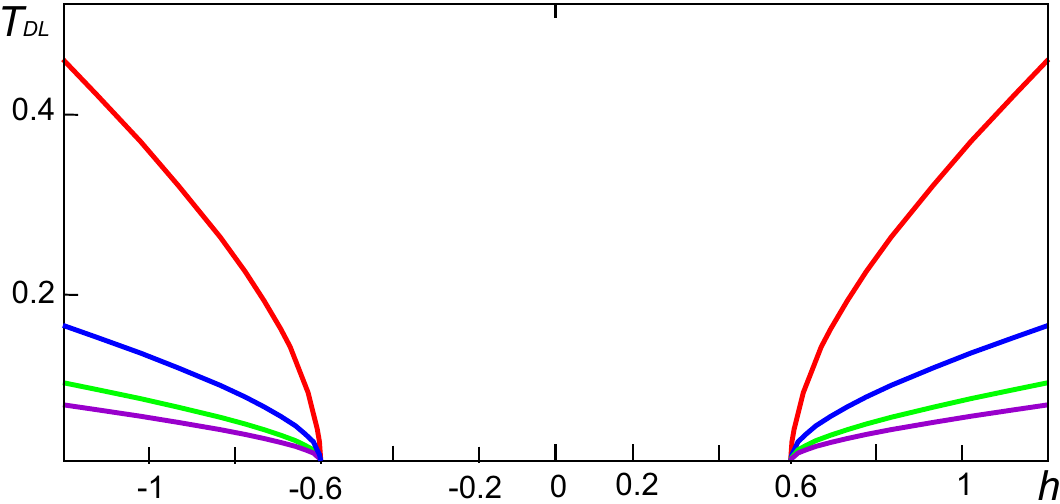}}
 \caption{The field dependence of $T_{\s DL,n}$ for $n=0,1,2,3$ (from top to bottom) for $\gamma=0.8$.
 The origin of all $T_{\s DL,n}$ lies at the value of $h=\pm \sqrt{1-\gamma^2}=\pm 0.6$.  }
 \label{TDLFig}
\end{figure}

To visualize the complicated surfaces of the disorder line solutions in the parametric space we present the plots in Figs.~\ref{PhaseDiag}
and \ref{TDLFig}. The magnetic field $h_{\s DL}\left(\gamma ,T,n\right)$ on the disorder lines with different $n$ in the $(\gamma ,h)$ plane
is obtained from $T_{\s DL,n} \left(\gamma ,h\right)=T$ as
\begin{equation}
\label{hDL}
  h_{\s DL} \left(\gamma ,T,n\right)=\pm \sqrt{1-\gamma ^{2} } \sqrt{1+\frac{\pi ^{2} \left(2n+1\right)^{2} T^{2} }{\gamma ^{2} } }
\end{equation}
These curves are plotted in Figs.~\ref{PhaseDiag} for two different temperatures and several values of $n$.

Another view on ``disorder surfaces" is given by their cross sections in the $(h,T)$-plane. The sheets of the DL temperatures $T_{\s DL,n} (\gamma ,h)$
corresponding to different $n$, all sprout from the same origin, i.e., the circle $h^2 +\gamma^2 =1$.
The DL temperatures as functions of the field for a fixed $\gamma$ are shown in Fig.~\ref{TDLFig}.
In the limit $T \to 0$ all disorder lines collapse in Fig.~\ref{PhaseDiag} onto a single circle $h^2 +\gamma^2 =1$, in agreement with the classical results \cite{McCoyII:1971}. Barouch and McCoy were the first to our knowledge to find the finite-temperature disorder line in the $XY$ chain, and in particular, they found the leading oscillating mode (with $n=0$) in the $zz$ spin correlation function at $T>0$.

\begin{figure}[h]
\centering{\includegraphics[width=6.5 cm]{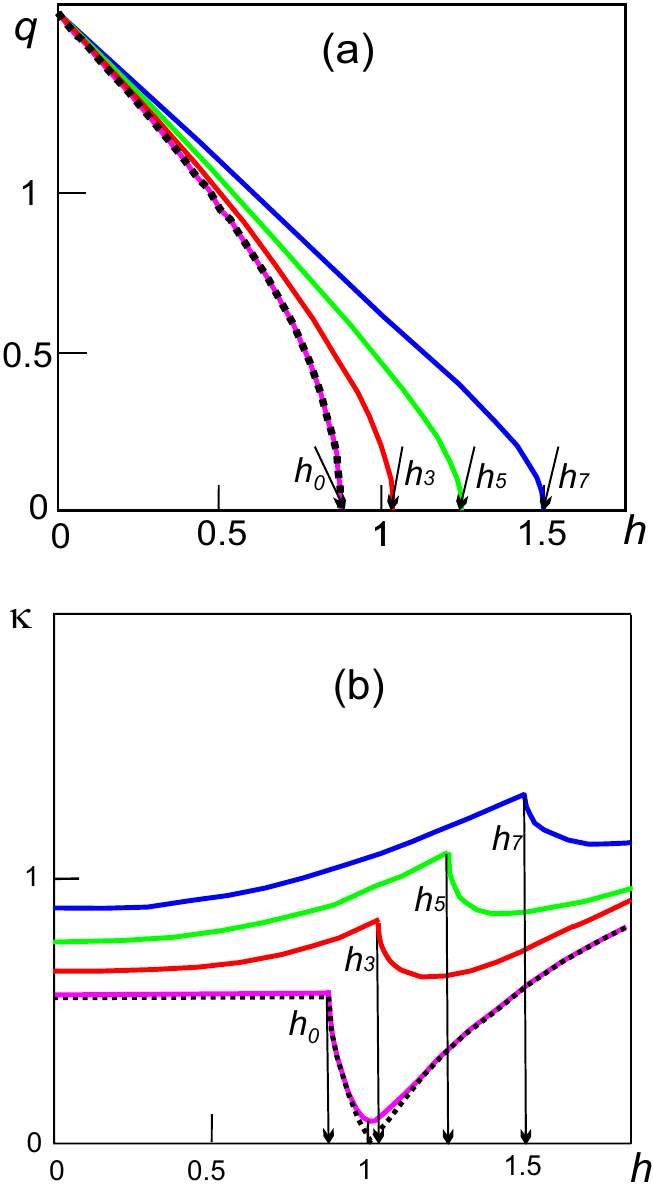}}
 \caption{The $h$-dependence of $q_n$ (a) and $\kappa_n$  (b) for $n=0$ (magenta), $n =3$ (red), $n=5$ (green), and $n=7$ (blue) at $\gamma=0.5$
 and $T = 0.015$. For brevity the critical fields $h_{\s DL}(n)$ are denoted as $h_n$. Dotted line is the $T \to 0$ limit. }
 \label{qkaphFig}
\end{figure}
\begin{figure}[h]
\centering{\includegraphics[width=6.5 cm]{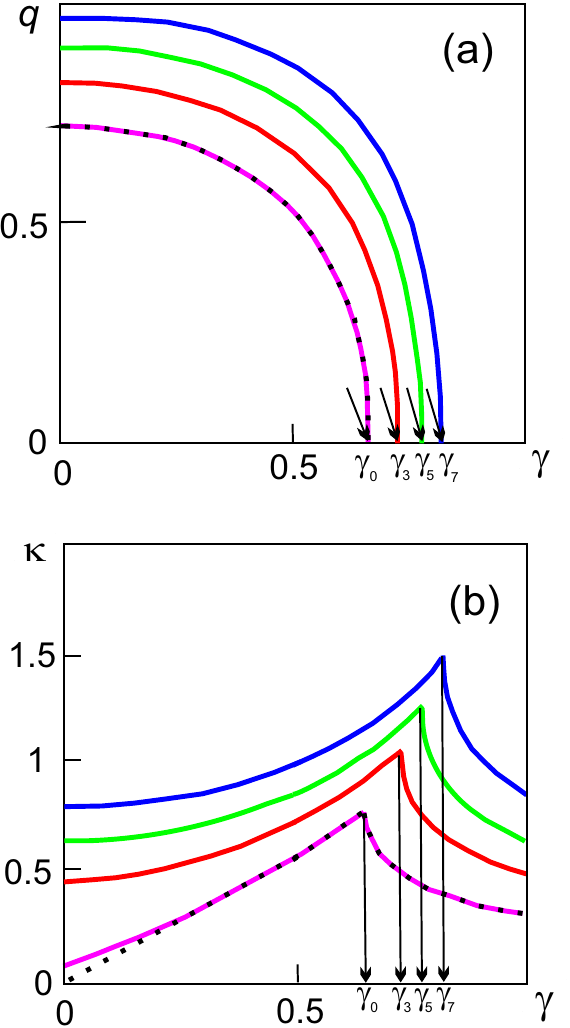}}
 \caption{The $\gamma$-dependence of  $q_n$ (a) and $\kappa_n$  (b)
 for $n=0$ (magenta), $n=3$ (red), $n=5$ (green), and $n=7$ (blue) at $h=0.75$ and $T = 0.015$.
 Dotted line is the $T \to 0$ limit.}
 \label{qkapgammaFig}
\end{figure}

In Fig.~\ref{qkaphFig} we plot the field dependencies of several wave numbers of oscillations $q_n$ and inverse characteristic length parameters  $\kappa_n$ of the oscillating modes. The minimal $\kappa_0$ can be identified with the inverse correlation length. As a result of level crossing there are cusps in $\kappa_n$ at the critical fields $h_{\s DL}(n)$, resembling similar features of corresponding quantities at the cascades of DLs found in the classical Ising chain \cite{Chitov:2017PRE}. In the zero temperature limit the inverse correlation length (dotted curve in Fig.~\ref{qkaphFig} (b)) vanishes at the quantum critical point $h=1$, as it must. Similar plots are presented in Fig.~\ref{qkapgammaFig} as functions of $\gamma$ at fixed value of the field.

One can easily find from Eq.\eqref{CosCompl} that the wave vectors of oscillations vanish when $T \to T_{\s DL,n}$ as
\begin{equation}
\label{qnTc}
  q_n \propto (T - T_{\s DL,n})^{\nu_{\s L}},
\end{equation}
while if the temperature is kept constant, $q_n$ shown in Fig.~\ref{qkaphFig}(a) vanish above the critical fields
$h_{\s DL} \left(\gamma ,T,n\right)$ \eqref{hDL} as
\begin{equation}
\label{qnhc}
  q_n \propto (h_{\s DL} - h)^{\nu_{\s L}}
\end{equation}
with the critical index of modulation $ \nu_{\s L}=1/2$ introduced earlier by Nussinov and co-workers \cite{Nussinov:2012}.

%
%
%
%
%
\subsection{Ground state factorization}\label{Factor}
%
%
%
%
%

The original idea by M\"{u}ller and coworkers \cite{Muller:1982,*Muller:1985} was to rotate each spin of the chain in
the $xz$ plane to make the transformed Hamiltonian ferromagnetic with the fully separable (factorized) ground state.
In the case of the $XY$ chain such factorizable doubly-degenerate ferromagnetic state occurs on the DL circle
$\gamma^2 +h^2 =1$ \cite{Franchini:2017}:
\begin{equation}
\label{GSpm}
  {\left| \Psi _{\pm}  \right\rangle} =\prod _{i=1}^{N}\left(\cos \theta~ {\left| \uparrow _{i}  \right\rangle} \mp \sin \theta~ {\left| \downarrow _{i}  \right\rangle} \right),
\end{equation}
with
\begin{equation}
\label{GSProd}
  \left\langle \Psi _{\pm } \Psi _{\pm } \right\rangle =1,~~
  \left\langle \Psi _{+} \Psi _{-} \right\rangle =\cos ^{N} 2\theta
\end{equation}
The angle of spin rotation $\theta$ is related to the roots \eqref{lampm} merging on the DL circle as:
\begin{equation}
\label{ThetaLam}
  \lambda _{+} =\lambda _{-} =\sqrt{\frac{1-\gamma }{1+\gamma } } =\cos 2\theta
\end{equation}
The problem of separable states in various spin models was quite vigorously studied in the literature,
see, e.g. \cite{Amico:2006,*Rossignoli:2009,*Illuminati:2009,*Cerezo:2017} and more references in there.
Our goal in this subsection is to present a consistent line of arguments relating the factorization of the ground state
to the analytical properties of the spectrum and, thus, of the generating function. To the best of our knowledge,
such analysis was not presented before.

The state \eqref{GSpm} is maximally disentangled, since the concurrence $\mathcal{C}$ introduced by Wootters \cite{Wootters:1998} as a measure of entanglement, vanishes on the DL circle. Indeed, the two-site concurrence can be calculated as
\begin{equation}
\label{C}
 \mathcal{C}=\sum _{m\ne n}{\left\langle \Psi  \right|} i\sigma _{m}^{y} i\sigma _{n}^{y} {\left| \Psi  \right\rangle} ~.
\end{equation}
The operator $\hat{P}_{n} =i\sigma _{n}^{y} $ of the rotation by the angle $\pi/2$ transforms a vector into the orthogonal one, so $\mathcal{C}=0$ in a factorized state. For the states \eqref{GSpm} one can easily verify:
\begin{equation}
\label{YY}
  {\left\langle \Psi _{+}  \right|} \sigma _{m}^{y} \sigma _{n}^{y} {\left| \Psi _{+}  \right\rangle} ={\left\langle \Psi _{-}  \right|} \sigma _{m}^{y} \sigma _{n}^{y} {\left| \Psi _{-}  \right\rangle} =0,~~\forall~ m \neq n.
\end{equation}

The constant correlation functions \cite{McCoyII:1971} on the DL circle is a hallmark of complete ground state factorization:
\begin{equation}
\label{XX}
\left\langle \sigma _{m}^{x} \sigma _{n}^{x} \right\rangle =
\left\langle \sigma _{m}^{x} \right\rangle \left\langle \sigma _{n}^{x} \right\rangle=
\sin ^{2} 2\theta =\frac{2\gamma }{1+\gamma },~~\forall~ m \neq n.
\end{equation}
We can trace such remarkable behavior of correlations from the analytical properties of spectrum $\varepsilon \left(z\right)$ and
closely related generating function $D(z)$. At $T=0$ the Majorana function \eqref{Loop} becomes
\begin{equation}
\label{GrT0}
  G_{n} = \oint _{\left|z\right|=1}\frac{dz}{2\pi i}  z^{n-1} D\left(z\right)~.
\end{equation}
\textit{The key property leading to the factorization result} is that the roots \eqref{ThetaLam} merge on the DL line  $\gamma^2 +h^2 =1$, and
the generating function \eqref{DzXY}
\begin{equation}
\label{DzDL}
  D(z)=\frac{z-\cos 2 \theta}{1-z \cos 2 \theta}
\end{equation}
becomes analytical $\forall~|z| \leq 1$. A straightforward calculation yields
\begin{eqnarray}
  G_n &=& 0, ~~n \geq 1 \\
  G_0 &=& - \cos 2\theta, \\
  G_{-n} &=& \sin^2 2\theta \cos^{n-1} 2\theta,~~ n \geq 1~.
\end{eqnarray}
The spin correlation functions $\mathcal{S}_{m-n}^{\alpha \alpha} \equiv \langle \sigma_n^\alpha \sigma_m^\alpha \rangle$
are given by the determinants of the Toeplitz matrices \cite{LiebSM:1961}. For the $yy$ components we trivially obtain
\begin{equation}
\label{Syy}
\mathcal{S}_{r}^{yy} =\det \left(\begin{array}{cccc} {0} & {G_{0} } & {G_{-1} } & {\begin{array}{cc} {...} & {G_{2-r} } \end{array}} \\ {0} & {0} & {G_{-2} } & {\begin{array}{cc} {...} & {G_{3-r} } \end{array}} \\ {0} & {0} & {0} & {\begin{array}{cc} {...} & {G_{4-r} } \end{array}} \\ {\begin{array}{c} {\vdots } \\ {0} \end{array}} & {\begin{array}{c} {\vdots } \\ {0} \end{array}} & {\begin{array}{c} {\vdots } \\ {0} \end{array}} & {\begin{array}{cc} {\begin{array}{c} {\ddots } \\ {...} \end{array}} & {\begin{array}{c} {\vdots } \\ {0} \end{array}} \end{array}} \end{array}\right)=0,
\end{equation}
in agreement with \eqref{YY}. The Toeplitz determinant for the $xx$ function
\begin{equation}
\label{Sxx}
\mathcal{S}_{r}^{xx} =\det \left(\begin{array}{cccc} {G_{-1} } & {G_{-2} } & {G_{-3} } & {\begin{array}{cc} {...} & {G_{-r} } \end{array}} \\ {G_{0} } & {G_{-1} } & {G_{-2} } & {\begin{array}{cc} {...} & {G_{1-r} } \end{array}} \\ {0} & {G_{0} } & {G_{-1} } & {\begin{array}{cc} {...} & {G_{2-r} } \end{array}} \\ {\begin{array}{c} {\vdots } \\ {0} \end{array}} & {\begin{array}{c} {\ddots } \\ {...} \end{array}} & {\begin{array}{c} {\ddots } \\ {0} \end{array}} & {\begin{array}{cc} {\begin{array}{c} {\ddots } \\ {G_{0} } \end{array}} & {\begin{array}{c} {\vdots } \\ {G_{-1} } \end{array}} \end{array}} \end{array}\right)
\end{equation}
is quite special: one can use the first row decomposition repeatedly to obtain
\begin{equation}
\label{Sxx1}
\mathcal{S}_{r}^{xx} = \sum _{n=1}^{r}\left(-G_{0} \right)^{n-1} G_{-n} \mathcal{S}_{r-n}^{xx},
\end{equation}
whence the result \eqref{XX} $\mathcal{S}_{r}^{xx}=\sin^2 2\theta$ for arbitrary $r$ can be proved by induction.

%
%
%
%
%
\section{2D Ising models}\label{Ising}
%
%
%
%
%
Due to correspondence between principal eigenvectors of transfer matrices of 2D Ising models and ground states of quantum chains
the above results can be applied to find disorder lines in the former.
This correspondence is stemming from commutation of the transfer matrix of a given Ising model (square, triangular, hexagonal, etc) with the
Hamiltonian of the quantum chain at particular values of model's couplings
\cite{Suzuki-I:1971,*Suzuki:1971,Krinsky:1972,Stephen:1972,Peschel:1982,Rujan-II:1982}.
From analysis of the eight-vertex model it is also possible to establish eqivalence of solvable 2D Ising models to free fermions \cite{FanWu:1970,WuLin:1987}.  In this section we extend the present analysis to apply it for Ising models which possess disorder lines.

Due to aforementioned correspondence between the classical and quantum models, the Gibbs thermal average of two Ising spins can be evaluated as a ground state average of the quantum spins \cite{Suzuki-I:1971,*Suzuki:1971,Krinsky:1972,Stephen:1972,Peschel:1982}. Thus the disorder lines analysed in the previous sections as points where oscillations of the correlation functions of the quantum model set in, are also points of oscillations of thermal correlation functions in the classical model.

The relations for the quantum-classical correspondence were given in detail in \cite{Stephen:1972} using the quantum cluster model
\begin{equation}
\label{tauHam}
   H =-\sum_{n=1}^{N} \Big\{ \frac{J}{4}  \big[ (1+\gamma )
 \tau_{n}^{x}
 - (1-\gamma )\tau_{n-1}^{z} \tau_{n}^{x} \tau_{n+1}^{z} \big]
 + \frac12 h \tau_{n}^{z} \tau_{n+1}^{z} \Big\}
\end{equation}
where $\tau$-s are also the Pauli matrices. The cluster Hamiltonian \eqref{tauHam} maps onto the $XY$ chain \eqref{XYHam} by the duality transformation
\cite{Peschel:1982,Peschel:2004}:
\begin{equation}
\label{tausigma}
  \tau_{n}^{x}=\sigma_{n-1}^{x} \sigma_{n}^{x}~,~~\tau_{n}^{z} \tau_{n+1}^{z}=\sigma_{n}^{z}~.
\end{equation}
The thermal average of two Ising spins $s_n$ is given by ground-state correlation functions of two $\tau$ spins or of the string of $\sigma$ spins:
\begin{equation}
\label{DualCorrFun}
 \langle s_{\s L} s_{\s R} \rangle_{\s \mathrm{Gibbs}} = \langle \tau_{\s L}^{z} \tau_{\s R}^{z} \rangle_{\s \mathrm{GS}}  =  \langle \prod_{l=L}^{R-1} \sigma_{l}^{z}   \rangle_{\s \mathrm{GS}}
\end{equation}
This is the way to recover the results of Stephenson \cite{Stephenson:1970PRB} for the frustrated triangular Ising model with $J_{1},J_{2},J_{3}$ couplings \cite{Peschel:1981,Peschel:1982}. For the transfer matrix along the direction of $J_{3}$ exchange
the correspondence between parameters of the quantum chan and the Ising model reads \cite{Stephen:1972,Peschel:1982}
\begin{eqnarray}
\label{Triangle}
 h&=& \frac{S_{1} S_{2} C_{3} +C_{1} C_{2} S_{3} }{C_{3} }~ ,  \nonumber \\
  \gamma &=& \frac{1}{C_{3}}  ~,  \\
  C_{i} &\equiv& \cosh 2\beta J_{i} ~,~   S_{i} \equiv \sinh 2\beta J_{i} \nonumber
\end{eqnarray}
The equations of Stephenson \cite{Stephenson:1970PRB} for $T_{\s DL}$ follow from the condition $h^2+\gamma^2=1$ expressed via Ising couplings
\eqref{Triangle}. The roots $\lambda_\pm$ \eqref{lampm} of the spectrum on the complex plane \eqref{kC} become complex conjugate on the disorder line,
with the wave number of oscillations
\begin{equation}
\label{qDL}
  \sin ^{2} q =\frac{1-\gamma^2- h^{2} }{1-\gamma^2}
\end{equation}
smoothly growing inside the oscillating phase in agreement with \eqref{qnhc}, while the inverse correlation length
\begin{equation}
\label{kapDL}
  \kappa =\min \left\{- \ln |\lambda_\pm | \right\}
\end{equation}
has a cusp at the DL temperature \cite{Stephenson:1970PRB}.

There are also so-called disorder lines of the second kind  \cite{Stephenson:1970PRB}, when the wave vector of modulations does not follow
\eqref{qnTc} or \eqref{qnhc}, but instead changes discontinuously (see also \cite{Nussinov:2011,Nussinov:2012}). Below we consider two Ising models
possessing such DLs with  $q$ changing from $q=0$ to $q=\pi$, and we connect the properties of DLs of the models with their Lee-Yang zeros.

\begin{figure}[h]
\centering{\includegraphics[width=4.0 cm]{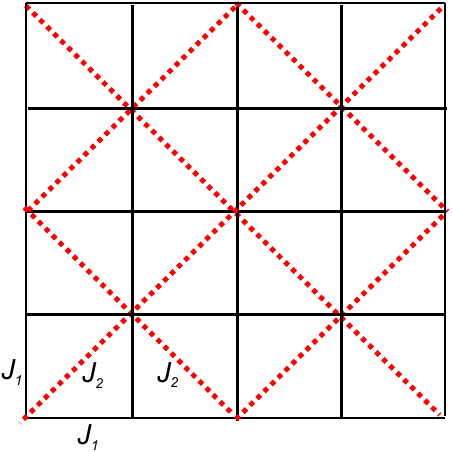}}
 \caption{The Union Jack model \cite{VaksLO:1966}.}
 \label{UJFig}
\end{figure}

\begin{figure}[h]
\centering{\includegraphics[width=6.0 cm]{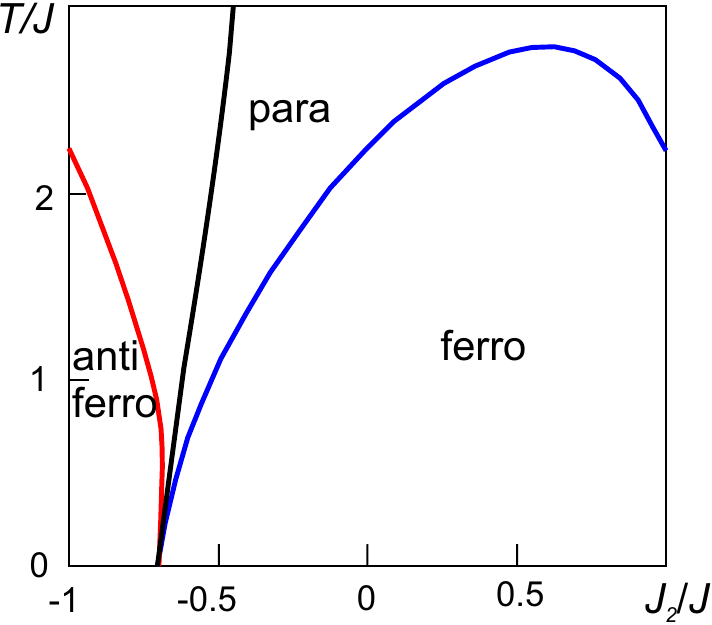}}
 \caption{Phase diagram of the Union Jack model \cite{VaksLO:1966} with $J\equiv \sqrt{J_{1}^{2} +J_{2}^{2}}$.
 The lines correspond to the transitions: red, PM-AF; blue, PM-FM; black, \textit{DL} of the second kind.}
 \label{VLOFig}
\end{figure}

One of these is the frustrated Ising model on the Union-Jack lattice first solved in \cite{VaksLO:1966}. The lattice with two couplings is shown in Fig.~\ref{UJFig}.
The model possesses the ferromagnetic (FM) and the antiferromagnetic (AF) phases separated by the paramagnetic (PM) one, for the
phase diagram, see Fig.~\ref{VLOFig}.
Stephenson found the DL of the second kind such that the PM phase is divided into two parts: the part adjacent to the AF phase has
oscillations with $q=\pi$ while the part neighboring the FM phase has $q=0$. \cite{Stephenson-II:1970,Stephenson:1970PRB,Peschel:1982}

Fan and Wu have shown \cite{FanWu:1970} that the Ising model on the Union-Jack lattice is equivalent to the eight-vertex model
in the free-fermionic limit which is:
\begin{equation}
\label{8VFF}
\omega _{1} \omega _{2} +\omega _{3} \omega _{4} =\omega _{5} \omega _{6} +\omega _{7} \omega _{8}.
\end{equation}
According to \cite{WuLin:1987}
\begin{eqnarray}
\label{8VIsing}
&~& \omega _{1} = 2e^{2K_{2} } \cosh 4K_{1} ,~  \omega _{2} =2e^{-2K_{2} } ,~  \omega _{3,4} =2, \nonumber \\
&~& \omega _{5,6,7,8} = 2\cosh 2K_{1} ,~  \mathrm{where}~ K_{n} \equiv J_{n} /T,
\end{eqnarray}
so the free-fermion condition \eqref{8VFF} yields
\begin{equation}
\label{FFUJ}
  2\omega _{5} {}^{2} =4+\omega _{1} \omega _{2}.
\end{equation}
For the frustrated model with $K_{1}>0$, $K_{2} <0$, and $-\left|K_{1} \right|<K_{2}$, when
\begin{equation}
\label{8VXY}
\gamma =\frac{\omega _{5} {}^{2} }{\omega _{1} +\omega _{2} } =\frac{4+\omega _{1} \omega _{2} }{2\left(\omega _{1} +\omega _{2} \right)} ,~
h=\frac{1}{4} \left(\omega _{1} -\omega _{2} \right),
\end{equation}
and $\cosh 4K_{1} >e^{-4K_{2} }$, the diagonal transfer  matrix of the Union-Jack or the free-fermion eight-vertex models
commutes with the ferromagnetic $XY$ Hamiltonian \eqref{XYHam},  while when $\cosh 4K_{1} <e^{-4K_{2} }$ it commutes with
the Hamiltonian \eqref{XYHam} for the case of antiferromagnetic coupling and negative field \cite{Krinsky:1972}.

\begin{figure}[h]
\centering{\includegraphics[width=6.0 cm]{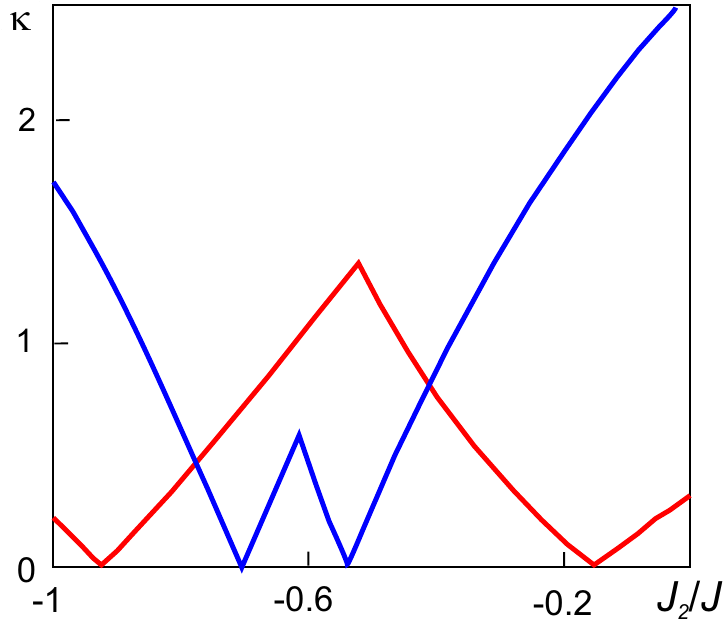}}
 \caption{Inverse correlation lengths in the Union Jack model
 for $T=2J$  (red) and $T=J$ (blue).
 The cusps on the DL of the second kind occur in the PM phase localized between to critical points
 of  PM-AF or PM-FM transitions where $\kappa=0$.}
 \label{VLOCorrFig}
\end{figure}

The thermal transition into the FM phase at $e^{-4K_{2} } +2e^{-2K_{2} } =\cosh 4K_{1} $ corresponds to the quantum transition in the chain \eqref{XYHam}
at $h=1$ and $\gamma= (\omega_2+2)/4$; the similar transition into the AF phase at $e^{-4K_{2} } -2e^{-2K_{2} } =\cosh 4K_{1} $
corresponds to the quantum transition in the AF chain at $h=-1$ and $\gamma= (\omega_2-2)/4$. Disorder line of the second kind
at $\cosh 4K_{1} =e^{-4K_{2} }$ corresponds to $h=0$ in the both ferromagnetic and antiferromagnetic chains and $\gamma =\left(4+\omega _{2}^{2} \right)/4\omega _{2} =\cosh K_{2} >1$. The phase diagram of the Union-Jack Ising model can be obtained from mappings \eqref{8VIsing},\eqref{FFUJ},\eqref{8VXY}. In particular, the cusp in the correlation length can be detected, as shown in Fig.~\ref{VLOCorrFig}.

\begin{figure}[h]
\centering{\includegraphics[width=4.0 cm]{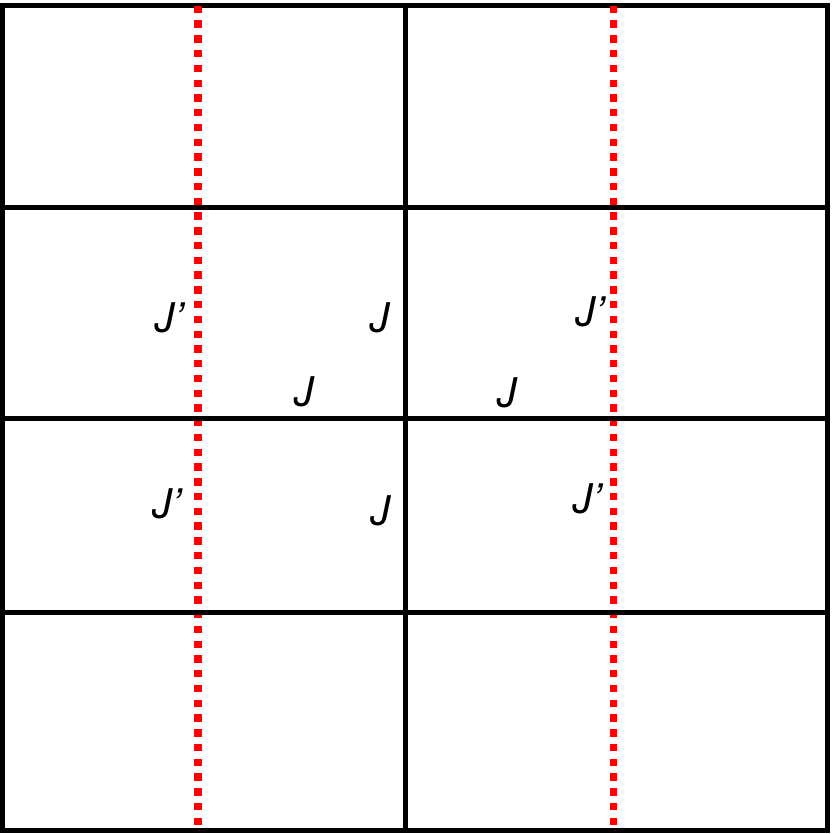}}
 \caption{Pile-up dominoes (PUD) model \cite{Andre:1979}. }
 \label{PUDFig}
\end{figure}

Similar situation occurs  in the piled-up dominoes (PUD) model considered in \cite{Andre:1979}. It is defined on the
lattice shown in  Fig.~\ref{PUDFig}.
The model has the transfer matrix $V$ in vertical direction, which can be represented as
\begin{equation}
\label{VPUD}
  V=A \exp \left( \mathcal{H}_{\mathrm{chain}} \right) ~,~~ \mathcal{H}_{\mathrm{chain}} = \sum _{k} \varepsilon (k)  c_{k}^\dag c_{k}~.
\end{equation}
\begin{widetext}
The fermionic spectrum is defined as
\begin{equation}
\label{epsPUD}
\cosh \varepsilon _{k} =\cosh 4K^{*} \cosh 2\left(K+K'\right)-\cos k\sinh 4K^{*} \sinh 2\left(K+K'\right)-2\sin 2K^{*} \sinh 2K'\sin ^{2} k~,
\end{equation}
in terms of the PUD couplings $K=J/T$, $K'=J'/T$.  $K^{*}$ is given by $\sinh 2K^{*} \sinh 2K=1$.
The model has PM-FM and PM-AF transitions when $\varepsilon(0)=0$ and  $\varepsilon(\pi)=0$, correspondingly \cite{Andre:1979}. See Fig.~\ref{PUDFig}.
More solutions for $\varepsilon(z)=0$ are found inside the circle $|z|=1$ on the complex plane. From \eqref{epsPUD} we get the equation
\begin{equation}
\label{PUDroots}
  2\sinh 2K^{*} \sinh 2K'\left(\frac{z+z^{-1} }{2} \right)^{2} -\left(\tau _{+}^{2} -\tau _{-}^{2} \right)\left(\frac{z+z^{-1} }{2} \right)+\tau _{+}^{2} +\tau _{-}^{2} -2\sin 2K^{*} \sinh 2K'=0~,
\end{equation}
which yields the roots
\begin{equation}
\label{PUDroots2}
\frac{z_{\pm } +z_{\pm }^{-1} }{2} =e^{iq_{\pm } } \cosh \kappa _{\pm } =\frac{\tau _{+}^{2} -\tau _{-}^{2} \pm \sqrt{D} }{4\sin 2K^{*} \sinh 2K'} .
\end{equation}
\end{widetext}

\begin{figure}[h]
\centering{\includegraphics[width=6.0 cm]{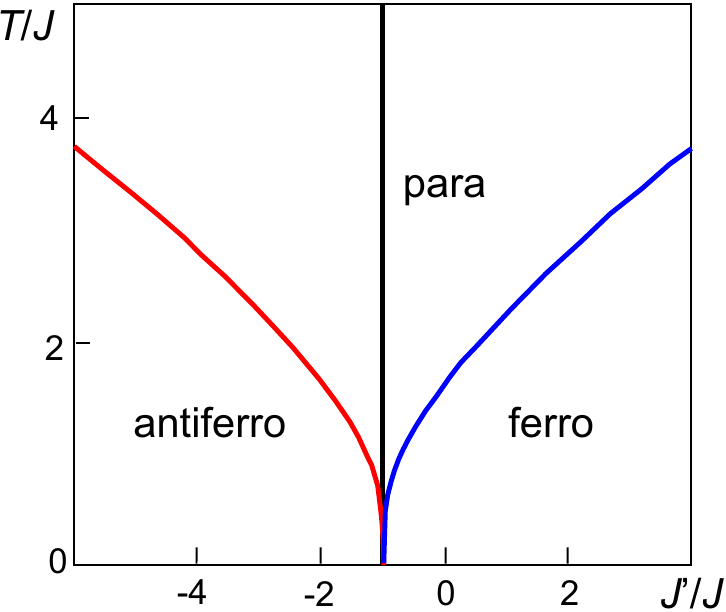}}
 \caption{Phase diagram of the PUD model \cite{Andre:1979}.
 The lines correspond to the transitions: red, PM-AF; blue, PM-FM; black, \textit{DL} of the second kind.}
 \label{DominoFig}
\end{figure}

\begin{figure}[h]
\centering{\includegraphics[width=6.0 cm]{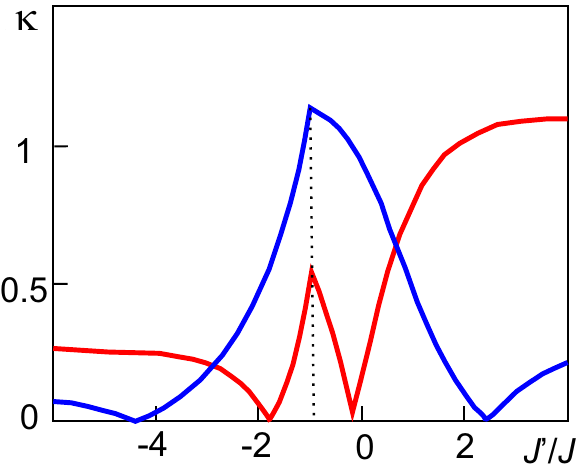}}
 \caption{Inverse correlation lengths in the PUD model for $T=1.5 J$ (red) and $T=3J$ (blue).
 The cusps on the DL of the second kind occur in the PM phase localized between to critical points
 of  PM-AF or PM-FM transitions where $\kappa=0$.}
 \label{PUDCorrFig}
\end{figure}

In the above equations we used the following notations:
\begin{eqnarray}
\label{Dtau}
  \tau _{\pm } &\equiv& \sinh \left(K+K'\pm 2K^{*} \right)~, \\
  D &\equiv& \left(\tau _{+}^{2} +\tau _{-}^{2} -4\sin 2K^{*} \sinh 2K'\right)^{2} -4\tau _{+}^{2} \tau _{-}^{2} ~.
\end{eqnarray}
The parametric curves of the FM and AF phase transitions \cite{Andre:1979} are recovered from the above equations at $\kappa _{+}=0$ and
$\kappa _{-}=0$, correspondingly, leading to
\begin{equation}
\label{FNAF}
  \tau _{\pm }^{2} =0~,~~\sinh 2K\sinh 2\left(K+K'\right)=\cos k=\pm 1~.
\end{equation}
In addition we find the DL of the second kind when
\begin{equation}
\label{PUDDL}
  \kappa _{+} =\kappa _{-}~~\mathrm{at}~~K+K'=0.
\end{equation}
To the best of our knowledge, this disorder line was not reported before. This feature inside the PM phase is accompanied by the cusps
of the correlation length, as shown in Fig.~\ref{PUDCorrFig}.

%
%
%
%
%
\section{2D/3D fermions}\label{3DFG}
%
%
%
%
%
For the grand canonical ensemble of non-interacting fermions the zeros of the partition function are readily found as \cite{LandauV5}:
\begin{equation}
\label{FreeF}
 \xi(\mathbf{k}) \equiv \varepsilon(\mathbf{k})- \mu = i \omega_n~,
\end{equation}
where $\mu$ is the chemical potential. They  are also zeros of the inverse single-particle temperature Green's function \cite{AGD:1963}:
\begin{equation}
\label{GF}
  G^{-1}(\mathbf{k}, \omega_n)=0~.
\end{equation}
In the limit $T=0$ equation \eqref{FreeF} becomes
\begin{equation}
\label{FS}
 \varepsilon(\mathbf{k})- \varepsilon_{\s F}= 0~,
\end{equation}
proving that the Fermi energy defines the surface of quantum criticality (gaplessness), and its appearance or restructuring constitutes a quantum phase transition \cite{Lifshitz:1960}.   This point has been pursued and elaborated by Volovik for quite a while \cite{Volovik:2003,Volovik:2007}, see also, e.g., \cite{Horava:2005}.

Equation for zeros of the partition function \eqref{FreeF} can be studied in the range of complex temperature or magnetic field. We follow our earlier
analysis and analytically continue the spectrum $\varepsilon(\mathbf{k})$ onto the complex plane as $k=q + i \kappa$.

The above equations are valid for any type of free fermionic Hamiltonian in two or three spatial dimensions, and analysis of \eqref{GF} can be done for tight binding lattice models, Dirac or topological materials, or even larger class of models, see \cite{Nussinov:2011,Nussinov:2012,Ogilvie:2020}.
For simplicity we choose to deal with the $3D$ non-relativistic gas of fermions with the spectrum $\varepsilon(\mathbf{k})=\frac{k^2}{2m}$.

The chemical potential of the degenerate ($T \ll \varepsilon_{\s F}$) Fermi gas to lowest order \cite{LandauV5}:
\begin{equation}
\label{mu}
   \frac{\mu}{\varepsilon_{\s F}} =1 -\frac{\pi^2}{12} \left(\frac{T}{\varepsilon_{\s F}}\right)^2 +\mathcal{O}(T^4)~,
\end{equation}
where $\mu(0)=\varepsilon_{\s F}=\frac{k^2_{\s F}}{2m}$.
To leading order the solutions of \eqref{FreeF} read
\begin{eqnarray}
  \label{qFF}
  \frac{q_n}{k_{\s F}} &\approx& 1+ \frac{\pi^2}{24} \left( 3 \left(2n+1\right)^2 -1 \right)
                     \left(\frac{T}{\varepsilon_{\s F}}\right)^2  \\
  \frac{\kappa_n}{k_{\s F}} &\approx&  \frac{\pi}{2}(2n+1)\frac{T}{\varepsilon_{\s F}}~,
  \label{kapFF}
\end{eqnarray}
and
\begin{equation}
\label{xiPoles}
   \xi(\mathbf{k}) - i \omega_n =\frac{1}{2m}(k-q_n+i \kappa_n)(k+q_n-i \kappa_n)
\end{equation}
The coordinate representation of the temperature Green's function \eqref{GF} is given by the following expression \cite{AGD:1963}
\begin{equation}
\label{Gr}
G(\mathbf{r}) =
\int \frac{d \mathbf{k}}{(2\pi)^3 }  e^{i\mathbf{kr}} n_{\s F} (\xi),
\end{equation}
where $n_{\s F} (\xi)$ is the Fermi-Dirac distribution function. It can be written as an expansion
similar to \eqref{GnSeries}:
\begin{equation}
\label{Gr3DSeries}
G(\mathbf{r}) =
\int_0^\infty \frac{k^2 d k}{(2\pi)^2 } \sin kr  \left( 1-
 \sum_{n=0}^\infty    \frac{4T \xi(k) }{\xi^2(k) +\omega_n^2} \right)~,
\end{equation}
where $\xi^2(k) +\omega_n^2$ in the above series can be easily factorized using the roots
\eqref{xiPoles} for zeros of the partition function. Integration by parts brings the above equation
to a better converging series:
\begin{eqnarray}
G(\mathbf{r}) = &-& \frac{1}{\pi^2 r^3} + \frac{2T}{\pi^2 m}
\int_0^\infty k dk \bigg[
\left( \frac{1}{r^3}- \frac{k^2}{2 r} \right) \cos kr \nonumber \\
&+& \frac{k}{r^2} \sin kr \bigg]
 \sum_{n=0}^\infty    \frac{\omega_n^2 -\xi^2(k)}{(\xi^2(k) +\omega_n^2)^2} ~,
\label{Gr3DSerBP}
\end{eqnarray}
Since \cite{Chitov:1998}
\begin{equation}
\label{Delta}
2T \sum_{n=0}^\infty    \frac{\omega_n^2 -\xi^2(k)}{(\xi^2(k) +\omega_n^2)^2}  \xrightarrow[T \to 0 ]{~} \delta(\xi)~,
\end{equation}
the zero-temperature $G(\mathbf{r})$ can be simply read off the integrand of \eqref{Gr3DSerBP}.
It oscillates with the wave number $k_{\s F}$, and
\begin{equation}
\label{GT0}
 G(\mathbf{r}) \xrightarrow[r k_{\s F} \gg 1 ]{~} -\frac{k^2_{\s F}}{2 \pi^2 r}\cos k_{\s F} r
\end{equation}
The integral \eqref{Gr3DSerBP} is quite cumbersome, but its key feature are the contributions from the poles \eqref{xiPoles}
with the leading term coming from $n=0$:
\begin{equation}
\label{Gas}
 G(\mathbf{r})  \sim  \frac{1}{r}\exp \left( \pm i q_{\s 0} r  -\kappa_{\s 0} r \right)
\end{equation}
The above result clarifies the physical meaning of zeros of the partition function with complex $k \in \mathbb{C}$:
the real part $q_{\s 0}(T)$ acts as a $T$-dependent Fermi wave vector which sets the period of spatial oscillations, while the imaginary part
determines the inverse correlation length $\kappa_{\s 0} =\pi T/2\varepsilon_{\s F}$.

It is also possible to introduce a finite-temperature generalization of the topological invariant  $N_1$ \cite{Volovik:2003} accounting for the 
$2 \pi N_1$ phase change of the Green's function \eqref{GF} while going around a path enclosing the Fermi surface. For the isotropic spectrum we parameterize the Fermi surface by the magnitude of the wave vector continued onto the complex plane $k \in \mathbb{C}$ with the poles  
$Q_n=[2m(\mu+i \omega_n)]^{1/2} \equiv q_n+i \kappa_n$, see  Eq.~\eqref{xiPoles}. Taking a small contour $C_0$ of radius $\epsilon$ around the zeroth pole $Q_0$:
\begin{equation}
\label{C0}
  C_0 :~ k=Q_0+\epsilon e^{i \varphi},~\varphi \in [0, 2\pi)~,
\end{equation}
the topological invariant is evaluated as a logarithmic residue at $Q_0$:
\begin{equation}
\label{N1}
  N_1=\oint_{C_0} \frac{dk }{2 \pi i} \partial_k \ln G^{-1}(k,\mu)=1~.
\end{equation}
The above definition smoothly evolves into the known result at zero temperature \cite{Volovik:2003,Volovik:2007}.

It is possible to define a similar topological invariant for the chain considered in Sec.~\ref{XYchain} using the loop integral around the poles $\Lambda_\pm$ (cf.  Eq.~\eqref{epsZ}) of the logarithmic derivative of $G^{-1}(k, \omega_n)=\varepsilon(k)-i \omega_n$. The latter is the temperature Green's function of the Bogoliubov fermions. However such topological number does not seem to be immediately useful. It is more relevant for various analyses \cite{Verresen:2018} to count the number of zeros of the partition function (or poles of the Green's function) inside the unit circle 
on the complex plane
\begin{equation}
\label{Nz}
    N_z=\oint _{\left|z\right|=1} \frac{dz}{2\pi i} \partial_z \ln \left(  \varepsilon^2 (z)+\omega_n^2   \right)~.
\end{equation}
Any change of $N_z$ means that a root (roots) crossed the unit circle $|z|=1$, which according to analysis of Sec.~\ref{XYchain}, signals a thermal or quantum phase transition. The definition \eqref{Nz} can be easily adapted for the tight-binding quadratic Hamiltonians in $d$-dimensions as well.

%
%
%
%
%
\section{Conclusion}\label{Concl}
%
%
%
%
%

An important motivation of this study is the conclusion of our earlier related work on the classical Ising chain \cite{Chitov:2017PRE}: the cascades of disorder lines in that model are zeros of the partition function with the complex magnetic field. Similarly, the appearance of modulations in the free fermion models stems from the analytical properties of zeros of their partition functions on the complex plane of the wave vectors $k \in \mathbb{C}$.

In this paper we propose a unifying framework based on the analysis of the roots for zeros of the partition function on the complex plane of wave numbers.
The general power of this approach is two-fold: first, it is not sensitive to the type of order parameter and can be used for both local and non-local parameters. Second, these roots combine all possible solutions corresponding to the continuous phase transitions, as well as to the disorder lines (or points of modulation transitions). We show how the analytical properties of the two-point Majorana correlation functions on the complex plane are related to the appearance of oscillations in those functions on the disorder lines and to the properties of the complex roots of the partition function.
In particular, even the known factorization of the ground state of the $XY$ chain on the disorder line, and consequently, vanishing entanglement, is shown
to follow directly from analyticity of the Majorana generating function inside the unit circle on the complex plane.

The disorder line transition is very weak, it is not straightforward to classify it in the standard scheme. For instance, for the $XY$ chain it was rigorously shown \cite{Maciazek:2016} that its ground state energy is smooth and even infinitely differentiable function on the disorder line.  The only nonanalytical clean-cut feature on the disorder line is a cusp in the behavior of the correlation length, which we explicitly calculated and plotted for the models considered. For the disorder lines with the modulation wave vectors continuously growing deep into the oscillating phase, cf. \eqref{qnTc} or \eqref{qnhc}, it is convenient to use the critical index of modulation $ \nu_{\s L}=1/2$ introduced earlier by Nussinov and co-workers \cite{Nussinov:2012}.

Most of our results are given for the simple quantum $XY$ chain in transverse field which is dually equivalent to free fermions.
We find an infinite cascade of disorder lines at finite temperature in this model and present results for such physical parameters as disorder line temperatures, correlation lengths, and wave vectors of oscillations. This was not analyzed before in the literature.
Since the transfer matrices of several 2D Ising models commute with the Hamiltonian of quantum chain at some special points \cite{Suzuki-I:1971,*Suzuki:1971,Krinsky:1972,Stephen:1972,Peschel:1981,Peschel:1982,Rujan-II:1982}, we used the results for the chain to detect the disorder lines in several frustrated 2D Ising models as well. The present formalism can be straightforwardly applied for tight binding lattice fermions or Fermi gas in two and three spatial dimensions. In particular, we find the complex roots for the zeros of the partition function of the 3D non-relativistic degenerate gas of fermions. The real part of this root is used to define the temperature-dependent Fermi wave vector which sets the period of spatial oscillations, while its imaginary part determines the inverse correlation length (gap) $\kappa \propto T$. The limit $T \to 0$ naturally leads to the definition of the Fermi energy as the surface of quantum criticality (gaplessness).

The appearance of modulation in correlation function seems to be a very common phenomenon. It occurs in a large variety of models  \cite{Nussinov:2011,Nussinov:2012,Salinas:2012}, including the recently reported pattern formation in the scalar Eucledian quantum field theory with a complex action \cite{Ogilvie:2020}. The important point to stress is that all these modulation transitions can be directly related to the partition function zeros, as done in the present study.

An interesting direction for the future work is to apply this formalism for the $XY$ chain with spatial and field modulations \cite{Chitov:2019},
including the interacting $XYZ$ case \cite{Chitov:2020}, and to analyze their Majorana edge states \cite{Karevski:2000,Chitov:2018,Vishvesh:2016};
to probe disorder lines in the deformed integrable Kitaev chains \cite{Chitov:2018} and in the (judiciously fermionized) Kitaev ladder or hexagonal models \cite{Xiang:2007,NussinovChen:2008}.

%
\begin{acknowledgments}
We thank Z. Nussinov for helpful discussions.
We acknowledge financial support from the Laurentian University Research Fund
(LURF) (G.Y.C.) and from the Ministry of Education and
Science of the Russian Federation (state assignment in the field of scientific
activity, project No.~0852-2020-0032 (BAS0110/20-3-08IF)) (P.N.T.).
\end{acknowledgments}
%


\bibliography{C:/Users/gchitov/Documents/Papers/BibRef/CondMattRefs}
%
\end{document}